%% file: main.tex
\documentclass[10pt,conference,letterpaper]{IEEEtran}
\IEEEoverridecommandlockouts

\usepackage{tikz}
\usepackage{algorithm}
\usepackage{algpseudocode}
\usepackage{cite}
\usepackage{amsmath,amssymb,amsfonts}
\usepackage{pifont}
\usepackage{graphicx}
\usepackage{booktabs}
\usepackage{textcomp}
\usepackage{mdframed}
\usepackage{hyperref}
\usepackage{listings}
\usepackage{xcolor} 
\def\BibTeX{{\rm B\kern-.05em{\sc i\kern-.025em b}\kern-.08em
    T\kern-.1667em\lower.7ex\hbox{E}\kern-.125emX}}

\renewcommand{\footnoterule}{%
  \kern -3pt                        
  \hrule width 0.4\columnwidth      
  \kern 2.6pt                       
}
\def\BibTeX{{\rm B\kern-.05em{\sc i\kern-.025em b}\kern-.08em
    T\kern-.1667em\lower.7ex\hbox{E}\kern-.125emX}}

\newcommand{\blackcircledwhite}[1]{%
  \tikz[baseline=(char.base)]{
    \node[shape=circle, fill=black, text=white, inner sep=1pt] (char) {\bfseries #1};
  }%
}

\newmdenv[
    backgroundcolor=gray!10,
    linecolor=gray!10,      
    shadow=true,
    shadowsize=2pt,
    shadowcolor=black!50,
    innertopmargin=5pt,
    innerbottommargin=5pt,
    innerleftmargin=5pt,
    innerrightmargin=5pt,
    linewidth=0pt,          
]{keytakeaway}

\makeatletter
\def\ps@IEEEtitlepagestyle{%
  \def\@oddfoot{\hbox{}\scriptsize
    \hfil
    \parbox{0.95\textwidth}{\centering
      \textcopyright~2025 IEEE. Personal use of this material is permitted.
      Permission from IEEE must be obtained for all other uses, in any current or future media,
      including reprinting/republishing this material for advertising or promotional purposes,
      creating new collective works, for resale or redistribution to servers or lists, or reuse
      of any copyrighted component of this work in other works.\\[2pt]
      DOI: \href{https://doi.org/10.1109/MASCOTS67699.2025.11283271}{10.1109/MASCOTS67699.2025.11283271}
    }
    \hfil}%
  \def\@evenfoot{}%
}
\makeatother
\begin{document}

\title{ Taming Cold Starts: Proactive Serverless Scheduling with Model Predictive Control

\thanks{This work was supported by the European Commission through the Horizon Europe project SovereignEdge.COGNIT (grant no. 101092711). Additional support was provided by the Wallenberg AI, Autonomous Systems and Software Program (WASP) funded by Knut and Alice Wallenberg Foundation.}
}



\author{
\IEEEauthorblockN{Chanh Nguyen\IEEEauthorrefmark{1}, Monowar Bhuyan\IEEEauthorrefmark{1}, Erik Elmroth\IEEEauthorrefmark{1}\IEEEauthorrefmark{2}}
\IEEEauthorblockA{\IEEEauthorrefmark{1}\it Department of Computing Science, Ume{\aa} University, SE-90187, Sweden}
\IEEEauthorblockA{\IEEEauthorrefmark{2}\it Elastisys AB, Ume{\aa}, Sweden}
\IEEEauthorblockA{Email: \{chanh, monowar, elmroth\}@cs.umu.se}
}
\maketitle
\begin{abstract}
\input{abstract}
\end{abstract}


\section{Introduction}

\input{intro}

\section{Related Work}
\label{sec:related}
\input{related2}

\section{Proactive Serverless
Scheduling with Model Predictive Control}
\label{sec:main_proposal}
\input{main_proposal}

\section{Experimental Setup}
\label{sec:experiment_setup}
\input{setup}

\section{Evaluation}
\label{sec:eva}
\input{evaluation}

\section{Conclusion and Future Work}
\input{conclusion}

\bibliographystyle{ieeetr}
\bibliography{acmart}

\end{document}

%% file: abstract.tex
Serverless computing has transformed cloud application deployment by introducing a fine-grained, event-driven execution model that abstracts away infrastructure management. Its on-demand nature makes it especially appealing for latency-sensitive and bursty workloads. However, the cold start problem, i.e., where the platform incurs significant delay when provisioning new containers, remains the Achilles' heel of such platforms. 

This paper presents a predictive serverless scheduling framework based on Model Predictive Control to proactively mitigate cold starts, thereby improving end-to-end response time. By forecasting future invocations, the controller jointly optimizes container prewarming and request dispatching, improving latency while minimizing resource overhead.

We implement our approach on Apache OpenWhisk, deployed on a Kubernetes-based testbed. Experimental results using real-world function traces and synthetic workloads demonstrate that our method significantly outperforms state-of-the-art baselines, achieving up to 85\% lower tail latency and a 34\% reduction in resource usage.

%% file: intro.tex
\noindent\textbf{Background.} Serverless computing~\cite{aslanpour2021serverless, jonas2019cloud} is a cloud execution model that provides \textit{Function-as-a-Service (FaaS)} capabilities, abstracting away infrastructure management while enabling fine-grained, event-driven function execution. While traditional Infrastructure-as-a-Service (IaaS) and Platform-as-a-Service (PaaS) models offer varying degrees of automation in provisioning and scaling, serverless platforms go further by providing per-request automatic provisioning, transparent scaling, and fine-grained billing based on actual execution time. This advantage eliminates the need for developers to manage runtime environments, instance lifecycles, or idle resource allocation.
Today, serverless platforms are increasingly adopted across a variety of domains, including web applications, data processing pipelines, IoT workloads, and machine learning inference~\cite{hong2024optimus, merlino2024faas}. This growth has been supported by the availability of both commercial platforms, such as AWS Lambda\footnote{\url{https://aws.amazon.com/lambda/}}, Google Cloud Functions\footnote{\url{https://cloud.google.com/functions}}, and Azure Functions\footnote{\url{https://azure.microsoft.com/en-us/services/functions}}, and open-source alternatives like Apache OpenWhisk\footnote{\url{https://openwhisk.apache.org/}} and OpenFaaS\footnote{\url{https://www.openfaas.com/}}, which offer developers greater flexibility and deployment control.

Despite its advantages, serverless computing suffers from \textbf{cold start latency}, i.e, the delay introduced when no warm function replica is available to serve a request. In such cases, the platform must initialize a new container, allocate resources, and load dependencies, leading to significantly higher response times compared to requests handled by already warm containers.
Such delays are especially harmful in \textit{latency-sensitive applications}, particularly those involving \textit{user-defined functions} with large dependencies (e.g., machine learning models)~\cite{sui2024pre}, where they can cause missed deadlines, failed requests, and significant QoS degradation, as illustrated in the real-world use case below.

\begin{figure}[H]
  \centering
  \includegraphics[width=0.55\columnwidth]{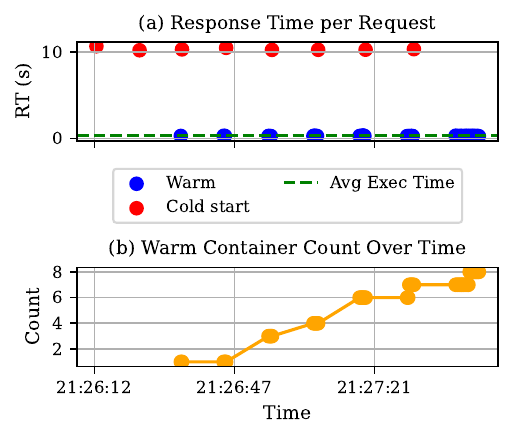}
  \caption{(a) Response time per request (in seconds). (b) Number of warm containers over time during 50 function invocations.}
  \label{fig:demo}
\end{figure}
\noindent\textbf{Real world example.} Consider an object detection function running the EfficientDet model~\cite{tan2020efficientdet} as part of a robotic application~\cite{nguyen2025tinykube}. The function is deployed on OpenWhisk, which runs on top of a Kubernetes cluster. Robots send frames captured by their cameras to the function to detect relevant objects and obstacles. Figure~\ref{fig:demo} shows the response time and the number of warm containers over time after sending 50 requests to the platform with randomly distributed arrival times. The average warm execution time is approximately 280 ms. However, during a cold start (eight cold start events are observed, highlighted in red,  resulting in eight warm containers by the end of the experiment) 
-- primarily due to the overhead of loading the TensorFlow runtime and the object detection model -- the response time for requests that triggered cold starts reaches approximately 10.5 seconds, corresponding to a 38$\times$ increase compared to the warm execution time. The earlier analysis~\cite{xiao2024making, bauer2024empirical} shows a similar finding, with cold start delays on platforms like AWS Lambda and Microsoft Azure reported to be 16$\times$--166$\times$ longer than the execution time.

\begin{figure}[H]
  \centering
  \includegraphics[width=0.7\columnwidth]{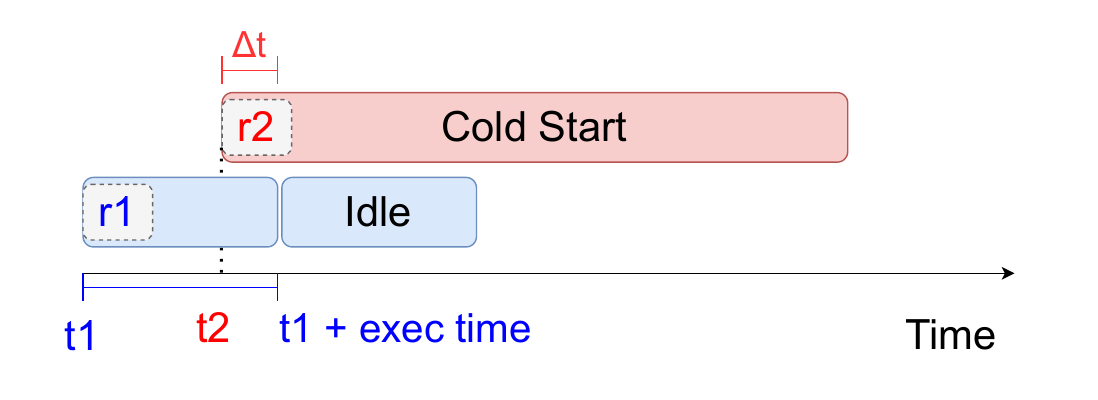}
  \caption{Unnecessary cold start due to lack of short-term request shaping.}
  \label{fig:event_Base}
\end{figure}

Most serverless platforms adopt a reactive, event-driven scheduling model that triggers a cold start as soon as no warm container is available, without deferring or batching incoming requests. While this strategy ensures responsiveness under load, it can lead to unnecessary delays. 
As illustrated in Figure~\ref{fig:event_Base}, consider a request $r_1$ arriving at time $t_1$ and assigned to an idle warm container, completing at $t_1 + \text{exec time}$. If a second request $r_2$ arrives at $t_2$ shortly before $r_1$ completes, and no other warm container is available, the platform launches a cold container for $r_2$, resulting in a significantly longer response time. 
However, if the system supports \textit{shaping incoming requests} to briefly wait for a soon-to-be-available warm container (i.e., $\Delta t = (t_1 + \text{exec time}) - t_2$), the cold start could be avoided, reducing the response time substantially.

\noindent\textbf{State of the art.} Existing efforts to reduce cold starts either shorten initialization time or limit their frequency. Cloud platforms like AWS Lambda and Azure Functions use static keep-alive windows (10--20 minutes)\cite{shahrad2020serverless}, which help under some workloads but are \textit{blind to invocation patterns, wasting resources}. Research efforts\cite{zhou2024tackling, li2022help, pan2022retention, xiao2024making} propose runtime-level optimizations like container reuse and sharing, but often raise deployment challenges due to \textit{security and isolation concerns}. Others apply reactive or predictive prewarming~\cite{wang2021lass, roy2022icebreaker}, yet often \textit{lack coordination with request admission}, causing delays under bursty workloads. We revisit these approaches in Section~\ref{sec:related}.

\noindent\textbf{Key insights and contributions.}
We propose a novel Model Predictive Control (MPC) approach to mitigate cold start delays in serverless computing. MPC~\cite{garcia1989model, morari1999model} optimizes system behavior over a receding time horizon by solving a constrained optimization problem at each control step. It is particularly well-suited for this problem because it: (1) enables joint optimization over request admission, container provisioning, and reclamation; and (2) naturally incorporates resource constraints such as container pool size and service limits, avoiding overprovisioning and queue overload.

Our key insight is that cold start delays can be reduced not just through provisioning, but also through predictive shaping, i.e., selectively deferring requests when short delays allow them to hit warm containers, thereby improving overall response latency.

We model serverless scheduling as a predictive control loop. The MPC controller uses Fourier-based forecasting to anticipate incoming request rates and decides at each control step how many containers to prewarm, reclaim, or serve, balancing latency and resource efficiency.

\vspace{0.5em}
\noindent Our contributions are as follows:
\begin{itemize}
\item \textbf{Predictive shaping for cold start mitigation.}  We show that short deferrals of request dispatch when guided by predictive logic can significantly reduce cold starts without degrading responsiveness (Section~\ref{sec:eva}).
\item \textbf{An MPC-based serverless scheduler.} We formulate cold start mitigation as a constrained optimization problem and design an MPC controller that jointly manages provisioning, reclamation, and dispatch under forecasted load (Section~\ref{sec:main_proposal}).
\item \textbf{Practical deployment in Apache OpenWhisk.} We implement our controller as a middleware layer that shapes and routes invocation traffic in real time, requiring no changes to the OpenWhisk core (Sections~\ref{sec:experiment_setup} and~\ref{sec:eva}).
\item \textbf{Comprehensive evaluation on real and synthetic traces.} We benchmark our system using production and synthetic workloads, demonstrating consistent improvements in response latency and resource efficiency compared to state-of-the-art baselines (Section~\ref{sec:eva}).
\end{itemize}


%% file: related2.tex
Prior work has tackled the cold start problem by either reducing its duration or minimizing its occurrence, typically falling into two main categories:

First, runtime-level optimizations intervene at the resource management layer to reduce the need for repeated function initialization. Solutions in this area focus on enabling container reuse, retention, or inter-function sharing~\cite{zhou2024tackling, li2022help, pan2022retention, xiao2024making}. For example, Zhou et al.~\cite{zhou2024tackling} propose a multi-level container reuse strategy across functions with similar environments, employing deep reinforcement learning for optimal reuse. Similarly, Pagurus by Li et al.~\cite{li2022help} mitigates cold start latency through inter-function container sharing, repurposing idle containers into lightweight zygotes for rapid specialization. Complementary efforts like Pan et al.~\cite{pan2022retention} and Xiao et al.~\cite{xiao2024making} focus on retention-aware caching frameworks, particularly in edge computing, to jointly manage container caching, request distribution, and cost.

Second, scaling-based approaches aim to ensure warm container availability by managing function replicas either reactively, in response to current load fluctuations, or proactively, based on workload prediction~\cite{wang2021lass, roy2022icebreaker}. Wang et al. introduce LaSS~\cite{wang2021lass}, a serverless edge computing platform that uses a queuing theoretic model for reactive replica scaling to meet high-percentile latency SLOs. However, LaSS is fundamentally reactive, adjusting container counts at fixed intervals without incorporating direct forecasting or fine-grained request-level responsiveness. This can lead to suboptimal performance for bursty workloads, as requests arriving before scaling completes may still experience full cold start latency. Another notable work is IceBreaker by Roy et al.~\cite{roy2022icebreaker}, a serverless scheduling framework that reduces cold start latency and keep-alive costs by leveraging heterogeneous servers for function prewarming. IceBreaker employs a function invocation predictor to capture time-varying patterns and a utility-based placement strategy. Despite its predictive prewarming, IceBreaker does not coordinate prewarming completion with request dispatch, nor does it shape incoming requests to wait for warm containers. Consequently, requests arriving before a prewarmed container is truly ready still incur the full cold start latency. 

In this paper, we address these limitations by proposing a solution that leverages Model Predictive Control (MPC) to proactively schedule both container prewarming and request shaping, enabling faster adaptation to workload fluctuations and reducing the impact of cold starts. To forecast incoming request rates, we adopt a Fourier harmonic prediction method inspired by~\cite{roy2022icebreaker}, and apply statistical clipping to constrain the predicted values within a plausible operational range, preventing overreaction to transient outliers in workload dynamics.

%% file: main_proposal.tex
\begin{figure}[H]
  \centering
  \includegraphics[width=\columnwidth]{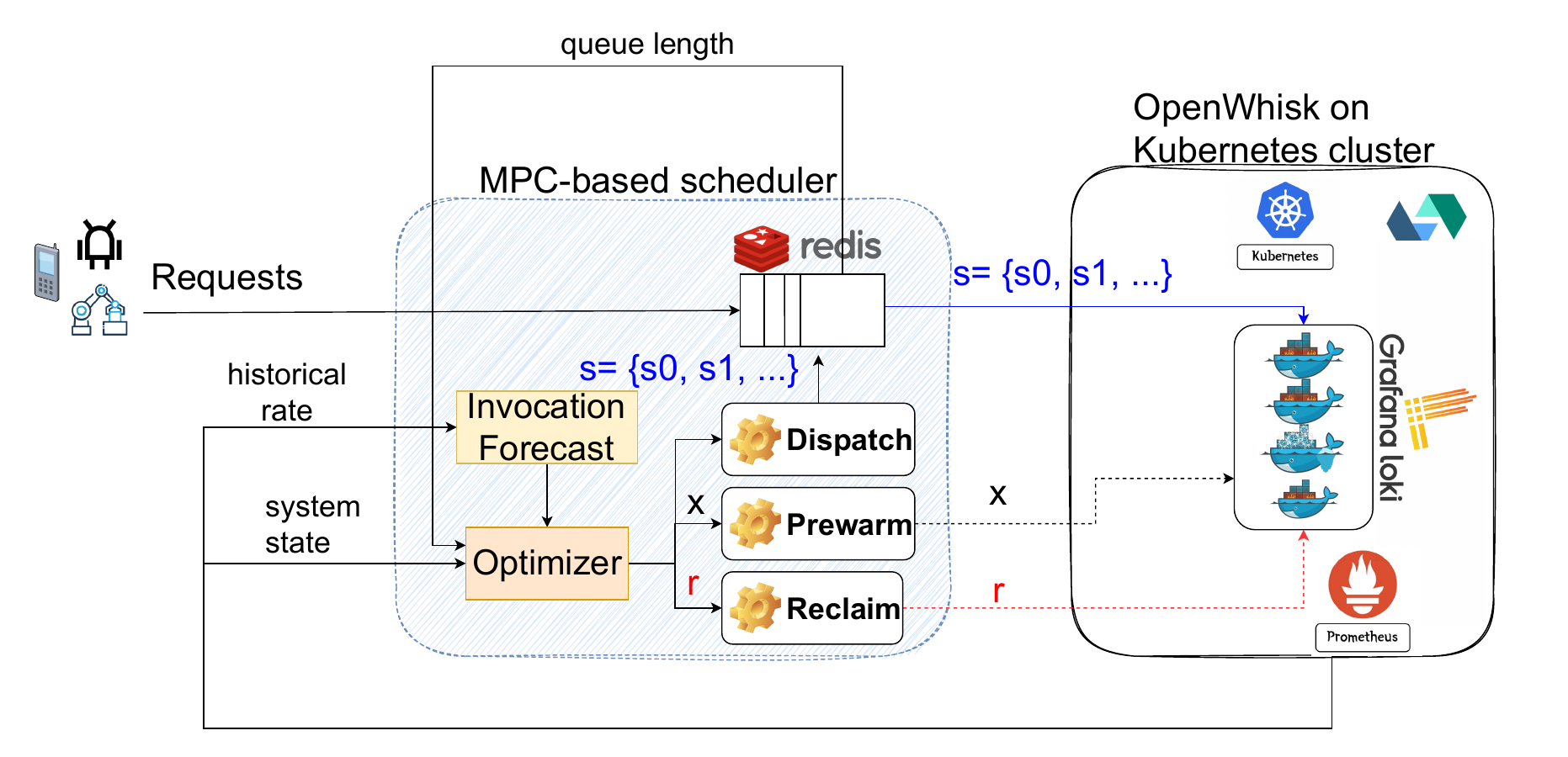}
  \caption{MPC-based proactive serverless scheduling architecture.}
  \label{fig:mpc}
\end{figure}

In this section, we present the architecture of the MPC-based proactive scheduler for serverless computing. Figure~\ref{fig:mpc} illustrates the main components and demonstrates how the scheduler interacts with an OpenWhisk deployment on a Kubernetes cluster. We assume that Prometheus\footnote{\url{https://prometheus.io/}} and Grafana Loki\footnote{\url{https://grafana.com/oss/loki/}} are deployed on the cluster: Prometheus is used to collect system metrics such as invocation rates and the number of active containers, while Loki is used to trace container behavior and determine when a container completes an activation.

In essence, the MPC scheduler is executed at every control interval $\Delta t$, following the sequence:
\blackcircledwhite{1} \textbf{Forecast future invocations}: the Invocation Forecast component uses historical metrics (e.g., invocation rate) from Prometheus  to predict the number of incoming requests over the next $H$ time steps;
\blackcircledwhite{2} \textbf{Optimization}: Based on the forecast, the MPC solves an optimization problem over the $H$-step horizon to determine how many containers to prewarm per time step, whether to reclaim idle containers, and how many requests to dispatch; and
\blackcircledwhite{3} \textbf{Execute current-step actions}: From the optimized plan, only the control actions for the current time step are executed via the corresponding actuators (dispatch, prewarm, reclaim), which interact with the OpenWhisk platform.
Below, we present the components involved in each process.

\subsection{Invocation Forecast}
At step~\blackcircledwhite{1} of the control cycle, the scheduler forecasts incoming function invocations over a prediction horizon of H time steps. Prior work~\cite{shahrad2020serverless, bauer2024globus} has shown that many serverless workloads exhibit periodic patterns that evolve over time. Traditional models such as histograms and ARIMA often struggle with such variability~\cite{roy2022icebreaker}. To address this, we adopt a Fourier-based extrapolation method~\cite{dyke2001introduction}, which captures multiple frequency components and offers greater robustness to shifting periodicity, enabling more accurate forecasting of both invocation timing and concurrency.
We emphasize that forecasting is not the central contribution of this work. Rather, we employ and extend the predictor proposed in~\cite{roy2022icebreaker} to better align with the needs of our scheduling framework.

The forecast at time step $t$ is given by:

\begin{equation}
\hat{\lambda}(t)= a t^2 + b t + c + \sum_{i=1}^{k} A_i \cos(2\pi f_i t + \phi_i)
\end{equation}

where \( a t^2 + b t + c \) represents the quadratic trend estimated from historical data; \( A_i \) and \( \phi_i \) denote the amplitude and phase of the \( i \)-th harmonic component; \( f_i \) is the corresponding frequency obtained from the discrete Fourier transform; and \( k \) is the number of harmonics used in the reconstruction.

While Fourier-based forecasts effectively capture periodic trends, they are unbounded and can return negative or overly large values, especially when trained on short or noisy histories~\cite{yang2024rethinking}. To improve robustness under non-stationary conditions, we apply \textit{statistical clipping}~\cite{fromming2021spectral} to constrain predictions within a realistic and safe operating range:

\begin{equation}
\hat{\lambda}_{\text{clipped}}(t) = \min \left( \max(0,\hat{\lambda}(t)), \mu + \gamma \cdot \sigma \right)
\end{equation}

Here, $\hat{\lambda}(t)$ denotes the raw forecast at time $t$, $\mu$ and $\sigma$ represent the mean and standard deviation of recent request rates, and $\gamma$ is a tunable confidence parameter. 
We implement the Fourier-based forecasting method in Python using standard scientific libraries, including NumPy for polynomial trend fitting and Fast Fourier Transform (FFT).
    
    
    
    
    
    
        
        
    

\subsection{Optimizer}
In step \blackcircledwhite{2}, the MPC controller solves an optimization problem over a horizon of \( H \) time steps to make scheduling decisions. We present below the individual cost components that define the objective function. For clarity, Table~\ref{tab:notation} summarizes the variables and parameters used in the MPC formulation.

\begin{table}[ht]
\centering
\caption{Summary of variables and parameters used in the MPC formulation}

\label{tab:notation}
\begin{tabular}{ll}
\toprule
\textbf{Symbol} & \textbf{Description} \\
\midrule
$q_k$ & Queue length at time step $k$ \\
$w_k$ & Number of warm containers at time step $k$ \\
$s_k$ & Number of requests served at time step $k$ \\
$x_k$ & Number of cold starts initiated at time step $k$ \\
$r_k$ & Number of containers reclaimed at time step $k$ \\
$\lambda_k$ & Number of incoming requests at time step $k$ \\
$\mu = 1 / L_{\text{warm}}$ & Service rate of a warm container \\
$w_{\max}$ & Maximum number of warm containers \\
$\text{readyCold}(k)$ & Cold-started containers ready at time $k$ \\
$L_{\text{warm}}$ & Warm container execution latency (s) \\
$L_{\text{cold}}$ & Cold start initialization latency (s) \\
$\alpha$ & Cost weight for cold start delay \\
$\beta$ & Cost weight for warm queue wait \\
$\gamma$ & Cost weight for overprovisioning \\
$\delta$ & Cost weight for initiating cold starts \\
$\eta$ & Reward weight for reclaiming containers \\
$\rho_1, \rho_2$ & Weights for provisioning smoothness \\
\bottomrule
\end{tabular}
\end{table}

\subsubsection{Cold delay penalty}

Let \( \text{ColdDelay}_k \) denote the latency penalty at time step \( k \) incurred when the number of available warm containers is insufficient to serve all incoming requests.

We define \( L_{\text{cold}} \) as the initialization latency of a cold container, and \( L_{\text{warm}} \) as the execution time in a warm container. Let \( \lambda_k \) represent the number of incoming requests at time step \( k \), and let \( w_k \) denote the number of warm containers available at that time. 
The cold delay penalty at time step \( k \) is computed as:
\begin{equation}
    \text{ColdDelay}_k = \alpha \cdot \max\left(0, \lambda_k - \mu \cdot w_k\right) \cdot \left(L_{\text{cold}} + L_{\text{warm}}\right)
\end{equation}
where \( \mu = \frac{1}{L_{\text{warm}}} \) is the service rate per warm container, and \( \alpha \geq 0 \) is a tunable cost weight that determines the importance of cold start penalties. Setting \( \alpha = 0 \) disables cold start awareness in the controller, while higher values encourage more aggressive prewarming to avoid latency.

\subsubsection{Queue waiting cost}
Let \( \text{WaitCost}_k \) denotes the accumulated delay experienced by queued requests, assuming each must wait for a warm container available. We estimate \( \text{WaitCost}_k \) as:
\begin{equation}
\text{WaitCost}_k = \beta \cdot q_k \cdot L_{\text{warm}}
\end{equation}
where \( \beta \geq 0 \) is a tunable cost weight that determines the penalty assigned to request queuing, and \( q_k \) is the number of queued requests at time step \( k \).
 
\subsubsection{Cold start cost} Let \( \text{ColdStartCost}_k \) denote the overhead of initializing new containers when no warm ones are available. This cost includes container allocation, function loading, runtime setup, and dependency initialization (e.g., ML model loading), making it significantly more expensive than warm execution.
To discourage frequent cold starts unless necessary, we introduce a tunable weight \( \delta \geq 0 \) that balances responsiveness and efficiency in the MPC objective.

Accordingly, the cold start cost at time step \( k \) is estimated as:
\begin{equation}
    \text{ColdStartCost}_k = \delta \cdot x_k
\end{equation}
where \( x_k \) is the number of cold starts initiated at time step \( k \).

\subsubsection{Overprovisioning penalty} We define \( \text{OverProvision}_k \) as a penalty term that discourages excessive allocation of warm containers at time step $k$:

\begin{equation}
\text{OverProvision}_k = \gamma \cdot \max\left(0, \mu \cdot w_k - \lambda_k \right)
\end{equation}

Here, \( \gamma \geq 0 \) is a tunable weight that penalizes unused capacity. Increasing \( \gamma \) encourages the controller to minimize overprovisioning, while a lower value allows more slack in warm container allocation to prioritize responsiveness.

\subsubsection{Reclaim Reward}
We define \( \text{ReclaimReward}_k \) as a reward that encourages the controller to reclaim unused warm containers, which is computed as:
\begin{equation}
\text{ReclaimReward}_k = -\eta \cdot r_k
\end{equation}
Here, \( \eta \geq 0 \) is a reward weight that incentivizes reclaiming idle containers, and \( r_k \) is the number of containers reclaimed at time step \( k \).

\subsubsection{Smoothness Penalty}
Finally, to avoid abrupt changes in provisioning, we define \( \text{Smoothness}_k \) as a penalty on fluctuations in cold start and warm container counts. This term aims to stabilize the system and prevent oscillatory behavior, and is calculated as:
\begin{equation}
\text{Smoothness}_k = \rho_1 \cdot (w_k - w_{k-1})^2 + \rho_2 \cdot (x_k - x_{k-1})^2
\end{equation}

Here, \( \rho_1 \) and \( \rho_2 \) are tunable weights that penalize variations in the number of warm containers \( w_k \) and cold starts \( x_k \), respectively.

Having defined the individual cost components, we now formulate the overall objective of the MPC as minimizing the total cost over a prediction horizon of \( H \) time steps. The objective balances response latency  and resource usage, while also discouraging frequent changes in the number of active containers to reduce management overhead and system instability:


{\small
\begin{align}
\min \sum_{k=0}^{H-1} (&\underbrace{\text{ColdDelay}_k + \text{WaitCost}_k}_{\text{response latency}} \notag\\
&+ \underbrace{\text{OverProvision}_k + \text{ColdStartCost}_k + \text{ReclaimReward}_k}_{\text{resource usage}} \notag\\
&+ \underbrace{\text{Smoothness}_k}_{\text{system stability}})
\end{align}
}

At each control step \( k \in \{0, 1, \ldots, H-1\} \), the MPC decides the number of cold starts \( x_k \), container reclaims \( r_k \), and requests to serve \( s_k \), while updating the system states: queue length \( q_k \) and number of warm containers \( w_k \). These decisions are subject to the following system dynamics and constraints:

\begin{alignat}{2}
q_{k+1} &= q_k + \lambda_k - s_k               &\quad& \text{(queue dynamics)} \\
w_{k+1} &= w_k + \text{readyCold}(k) - r_k     &\quad& \text{(warm container update)} \\
s_k     &\leq \min(q_k,\ \mu \cdot w_k)        &\quad& \text{(serving capacity)} \\
r_k     &\leq w_k                              &\quad& \text{(reclaim bound)} \\
0       &\leq x_k \leq w_{\max}                &\quad& \text{(cold start limits)} \\
0       &\leq r_k \leq w_k                     &\quad& \text{(reclaim limits)} \\
0       &\leq w_k \leq w_{\max}                &\quad& \text{(warm container limits)} \\
0       &\leq s_k,\ q_k                        &\quad& \text{(non-negativity)} \\
\phantom{q_{k+1}} & r_k \cdot x_k = 0           &\quad& \text{(mutual exclusivity)}
\end{alignat}

Here,

\[
\text{readyCold}(k) =
\begin{cases}
x_{k-D}, & \text{if } k \geq D \\
0, & \text{otherwise}
\end{cases}
\]
 models the number of cold-started containers that become available at time step \( k \), based on a discrete cold start delay of \( D \) steps, where \( D = \left\lfloor L_{\text{cold}} / \Delta t \right\rfloor \) and \( \Delta t \) is the MPC control interval.

To solve the MPC optimization problem, we use the \texttt{cvxpy} library~\cite{diamond2016cvxpy}, a Python-embedded modeling language for convex optimization.

\subsection{Actuators}

At step~\blackcircledwhite{3} in the control cycle, the actuators: \emph{dispatch}, \emph{prewarm}, and \emph{reclaim} serve as the operational interface between the MPC controller and the OpenWhisk platform. 
Given the MPC decisions at time step \( k \), specifically \( s_k \), and either \( x_k \) or \( r_k \) (which are mutually exclusive as per constraint~(18)), these actuators execute the selected actions on the platform.

\textbf{The \emph{prewarm} actuator}, exposed via the function \texttt{launchColdContainers}($x_k$), triggers container initialization by issuing $x_k$ parallel \texttt{wsk} CLI calls to the function (specified via \texttt{function\_name}) with a custom \texttt{forcePrewarm=true} parameter, as shown in Listing~\ref{lst:prewarm-cmd}. The serverless function's handler checks this flag and skips actual execution logic, enabling lightweight warmup.

{\scriptsize
\begin{lstlisting}[caption={Prewarm $\text{count} = x_k$ containers.}, 
label={lst:prewarm-cmd}]
cmd = f"seq {count} | xargs -P{parallelism} -I{{}} wsk -i action invoke \
    {function_name} --param forcePrewarm true"
result = subprocess.run(cmd, shell=True, check=True,
                        stdout=subprocess.PIPE, stderr=subprocess.PIPE)
\end{lstlisting}
}

\textbf{The \emph{dispatch} actuator}, implemented as \texttt{dispatchRequests}($s_k$), is described in Algorithm~\ref{alg:mpc-dispatch}. It sends $s_k$ requests in batches, based on the number of available warm containers $w_k$ (line 2-5). Requests are retrieved from a Redis queue\footnote{\url{https://redis.io/glossary/redis-queue/}} (line 3) and dispatched asynchronously to the OpenWhisk API endpoint (line 5).


\begin{algorithm}
\caption{Dispatch Requests}
\label{alg:mpc-dispatch}
\begin{algorithmic}[1] 
\Require Requests to send $s_k$
\Require Number of warm containers $w_k$ 
\While{$s_k > 0$}
    \State $B \gets \min(s_k, w_k)$ \Comment{Batch size for this round}
    \State $R \gets \text{next } B \text{ requests from queue}$ 
    \ForAll{$r \in R$ \textbf{in parallel}}
        \State \texttt{submitRequestAsync}($r$)
    \EndFor
    \State $s_k \gets s_k - |R|$
\EndWhile 
\end{algorithmic}
\end{algorithm}



        

\begin{algorithm}
\caption{Reclaim Idle Function Containers}
\label{alg:reclaim-pods}
\begin{algorithmic}[1]
    \State $P \gets \text{\texttt{rankPods}}(r_k)$
    \If{$P = \emptyset$}
        \State \textbf{exit:} no container available
    \EndIf
    \State $L \gets \text{\texttt{listRunningFunctionPods}}()$
    \State $S \gets \{p \in P \mid p \notin L\}$ \Comment{Safe to reclaim}
    \ForAll{$p \in S$}
        \State \text{\texttt{drainAndReclaimPod}}($p$)
    \EndFor
\end{algorithmic}
\end{algorithm}

\textbf{The \emph{reclaim} actuator}, exposed via the function \texttt{reclaimIdleContainers}($r_k$), must ensure that no in-flight activations are disrupted during reclamation. 
The full logic is shown in Algorithm~\ref{alg:reclaim-pods}.
To guarantee safety, it restricts reclamation to containers confirmed to be idle: containers are ranked using a composite score that prioritizes \textit{low CPU/memory usage and long idle duration}, which helps reduce churn and prevent thrashing.

From this ranking, the top $r_k$ containers are selected as candidates (line 1). Before reclaiming, the system verifies that each candidate is no longer processing requests by querying the Loki log aggregation system. Specifically, it checks for the log message \texttt{[MessagingActiveAck] posted completion of activation} to confirm that the container has completed all assigned in-flight activations (line 5-6). Only containers that pass this check are considered safe and are reclaimed accordingly (lines 7–9).

%% file: setup.tex
\noindent\textbf{Experimental Platform.} 
 We deploy a single-node Kubernetes cluster using k3s on a server running Ubuntu 20.04, equipped with an AMD Opteron 6272 CPU (32 vCPUs at up to 2.1\,GHz) and 48\,GB of RAM. The node serves as both control plane and worker, with OpenWhisk installed to support serverless execution. A monitoring stack comprising Prometheus and Grafana Loki is also deployed to expose APIs for collecting metrics and logs from the MPC controller.
As our contribution centers on scheduling logic and request shaping rather than container placement or distributed orchestration, a single-node deployment is sufficient to validate the proposed control strategy.

To avoid resource contention and ensure stable, interference-free evaluation of scheduling decisions, we deploy the MPC scheduler and request generator on a separate local machine (Ubuntu 20.04, Intel Core i5-13500, 32\,GB RAM) on the same network as the Kubenertes cluster.


\noindent\textbf{Function.} We implement a serverless function image for object detection using EfficientDet~\cite{tan2020efficientdet}. On cold start, the function initializes the TensorFlow library, which dominates the startup latency. 
Profiling shows that warm executions average \( L_{\text{warm}} = 280\,\text{ms} \), while the initialization time due to cold starts is approximately \( L_{\text{cold}} = 10.5\,\text{seconds} \).

Each replica is limited to 256\,MB of memory and an estimated CPU usage of 0.5\,vCPU. Under these constraints, the serverless platform can support up to 64 concurrent replicas, bounded by CPU resources.

\noindent\textbf{Workload.} We implement a configurable workload generator to produce invocation requests to the serverless platform. The generator allows control over both the request arrival rate and the number of requests sent concurrently.
To create interl-arrival rate, we use: 
\begin{itemize}
    \item \textbf{Azure Function Traces.} We extract inter-arrival times from real-world invocation logs collected over two weeks from Microsoft Azure Functions~\cite{shahrad2020serverless}.

    \item \textbf{Synthetic Workload.} We synthesize inter-arrival patterns by randomly sampling burst durations (1–5)~s, idle periods (50–800)~s, and request rates (5–300)~req/s.
\end{itemize}

\noindent\textbf{Baseline Approaches.} We compare our MPC scheduler against the following baseline strategies:

\begin{itemize}
    \item \textbf{OpenWhisk Default Policy.} By default, OpenWhisk triggers a cold start when no warm container is available to handle an invocation. It keeps function containers in a warm state for up to 10 minutes after their most recent use.
\item \textbf{IceBreaker}~\cite{roy2022icebreaker} reduces cold starts and keep-alive costs through proactive prewarming and predictive function placement. It employs Fourier-based forecasting to estimate future invocations. Its key innovation lies in leveraging server heterogeneity: functions are initially placed on low-end servers for extended warm retention and later migrated to high-end servers as invocation likelihood increases. Since our evaluation assumes a single-server setup, we adapt IceBreaker to a homogeneous environment by disabling server-type–specific placements.

\end{itemize}

\noindent\textbf{Evaluation Metrics.}
We evaluate the proposed approach and the baselines using the following metric:  
\begin{itemize}
   \item \textbf{Total response time per request}: the end-to-end latency observed by the user, defined as the sum of queueing delay, cold start time, and execution time.
   \item \textbf{Resource efficiency}: measured by the number of containers and total keep-alive duration, reflecting the cost of maintaining warm containers to handle incoming requests.

\end{itemize}

%% file: evaluation.tex
\subsection{Prediction accuracy}
\begin{figure}[t]
\centering
\begin{tabular}{cc}
   \includegraphics[width=0.5\linewidth]{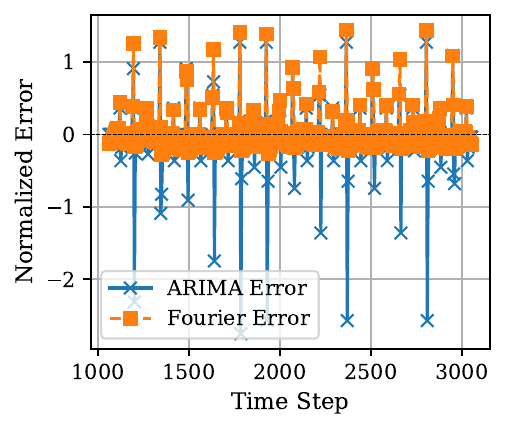} &
   \includegraphics[width=0.5\linewidth]{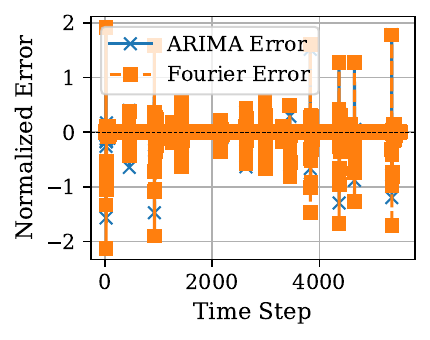} \\
   \small (a) Microsoft Azure Function & \small (b) Synthetic data \\
\end{tabular}
\caption{Forecast error of Fourier and ARIMA models in two experiments: (a) with Microsoft Azure Functions and (b) with synthetic data.}
\label{fig:forecast_error}
\end{figure}
Before evaluating the full control loop, we isolate the forecasting component to assess its accuracy, as accurate prediction of request arrival rates is critical for enabling proactive and effective scheduling decisions. To this end, we evaluate the predicted versus actual arrival rates in two experimental scenarios: using real-world traces from Azure Functions, and using synthetically generated arrival patterns. We implement the ARIMA time series model as a baseline for comparison with the Fourier-based forecasting approach. Figure~\ref{fig:forecast_error} presents the forecast errors of the two methods. In both experiments, the Fourier-based predictor outperforms ARIMA. Specifically, on the Azure Function dataset, Fourier achieves an accuracy of 86.2\%, compared to 82.5\% for ARIMA. On the synthetic dataset, both methods achieve comparable accuracy (ARIMA: 95.9\%, Fourier: 95.3\%). 
Notably, the runtime of the Fourier predictor (0.1\,ms) is over 100× faster than ARIMA (10\,ms) when performing rolling updates and prediction.

\subsection{Total response time per request}
\begin{figure}[t]
\centering
\begin{tabular}{cc}
   \includegraphics[width=0.5\linewidth]{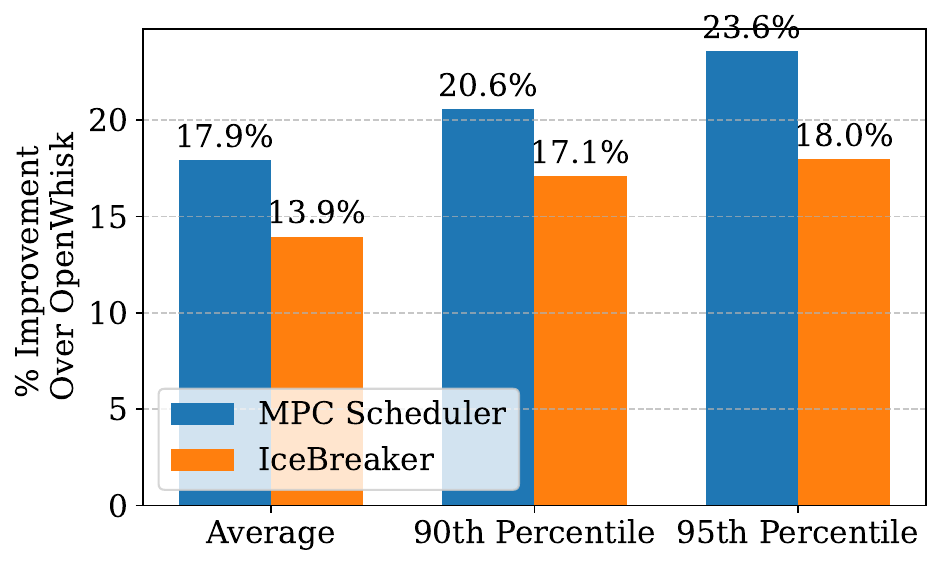} &
   \includegraphics[width=0.5\linewidth]{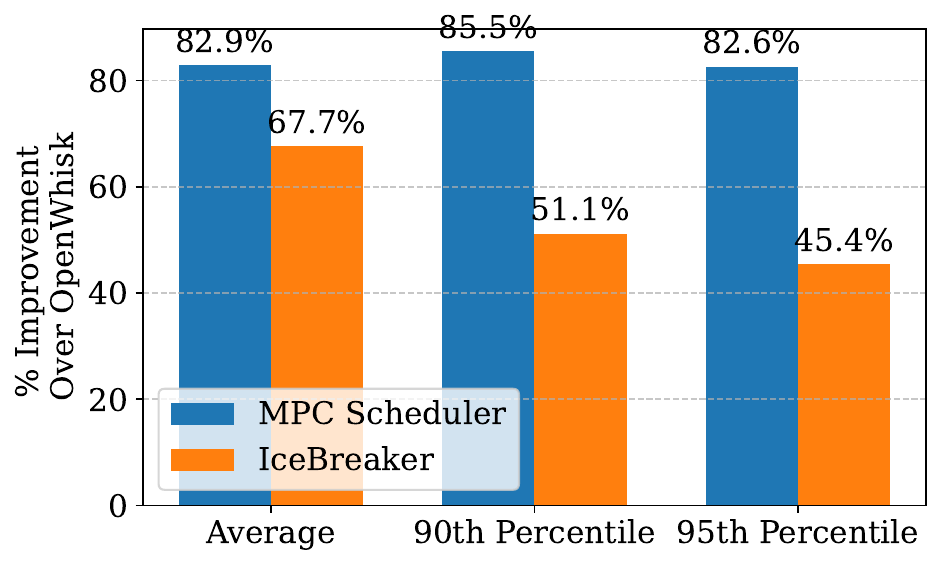} \\
   \small (a) Microsoft Azure Function & \small (b) Synthetic data \\
\end{tabular}
\caption{Percentage improvement in total response time (average, 90th, and 95th percentiles) over OpenWhisk. (a)  with Microsoft Azure Functions; (b) with synthetic data.}

\label{fig:cdf_response}
\end{figure}

All three approaches are evaluated under the same arrival patterns over a 60-minute period, using both the Azure Function trace and synthetic workloads. Each experiment begins with no warm containers available on the OpenWhisk platform.
Figure~\ref{fig:cdf_response} shows the percentage improvement in end-to-end response time over OpenWhisk’s default policy for the two approaches: MPC-Scheduler and IceBreaker.

The high accuracy of the Fourier-based forecasting enables both IceBreaker and MPC-Scheduler to proactively prewarm an appropriate number of function replicas, effectively mitigating cold start overhead in both the Azure trace and synthetic workload experiments. 
In the experiment using the Azure Function trace, the extracted inter-arrival rates exhibit steady, non-bursty behavior, resulting in limited improvement in response time compared to the default OpenWhisk policy. Specifically, MPC-Scheduler achieves a 17.9\% reduction in mean response time, while the improvements in the 90th and 95th percentile tail latencies are more pronounced at 20.6\% and 23.6\%, respectively.
The IceBreaker approach also shows improvements in end-to-end response time compared to the default OpenWhisk policy, with reductions of 13.9\% in average response time, 17.1\% in the 90th percentile, and 18\% in the 95th percentile tail latencies.

Under the synthetic workload, where request arrivals exhibit pronounced burstiness, i.e., many invocations occur within short time intervals, the improvements in response time over the default OpenWhisk policy are substantially greater for both MPC-Scheduler and IceBreaker. MPC-Scheduler achieves reductions of 82.9\%, 85.5\%, and 82.6\% in average, 90th percentile, and 95th percentile response times, respectively. IceBreaker also improves performance, with corresponding reductions of 67.7\%, 51.1\%, and 45.4\%.

In both experiments, MPC-scheduler consistently outperforms IceBreaker, despite both approaches employing the same underlying Fourier-based forecasting technique. This improvement is largely attributed to MPC’s joint optimization of request dispatching and container prewarming. By briefly queuing requests and aligning them with available warm containers, MPC-Scheduler effectively avoids unnecessary cold starts. In contrast, IceBreaker immediately forwards incoming requests to OpenWhisk, which can result in cold starts if no warm containers are available, even when forecasts are accurate.

\textit{These observations highlight that short deferrals of request dispatch, when guided by predictive logic, can significantly reduce the impact of cold starts without compromising responsiveness.}
\subsection{Resource usage and Keep-alive cost}
\begin{figure}[t]
\centering
\begin{tabular}{cc}
   \includegraphics[width=0.5\linewidth]{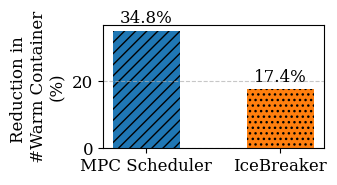} &
   \includegraphics[width=0.5\linewidth]{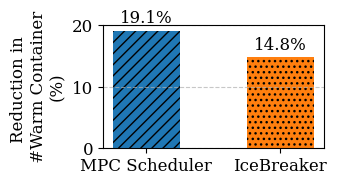} \\
a) Microsoft Azure Function  & b)  Synthetic data 
\end{tabular}
\caption{Percentage reduction in total warm container usage by MPC-Scheduler and IceBreaker compared to OpenWhisk’s default policy, measured at 1-minute intervals. (a)  with Microsoft Azure Functions; (b) with synthetic data.}

\label{fig:container}
\end{figure}

\begin{figure}[t]
\centering
\begin{tabular}{cc}
   \includegraphics[width=0.5\linewidth]{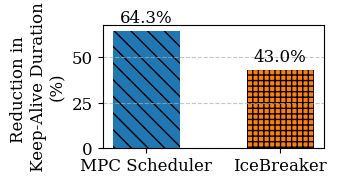} &
   \includegraphics[width=0.5\linewidth]{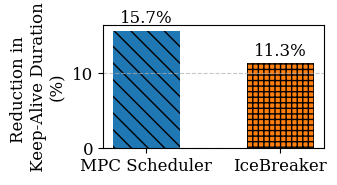} \\
a) Microsoft Azure Function  & b)  Synthetic data 
\end{tabular}
\caption{Percentage reduction in keep-alive duration achieved by MPC-Scheduler and IceBreaker, relative to OpenWhisk’s default policy.}

\label{fig:keep_alive}
\end{figure}

We collect the number of warm containers used by all three approaches at 1-minute intervals throughout the experiments. To quantify the relative change in resource usage, we compute the percentage difference in warm container count at each time step.  Additionally, for each container invoked by the serverless platform, we track the duration from its last activation until reclamation to evaluate the effective keep-alive time.

Figure~\ref{fig:container} presents the reduction in warm container usage achieved by MPC-Scheduler and IceBreaker, relative to the default OpenWhisk policy. Figure~\ref{fig:keep_alive} further illustrates the corresponding reduction in keep-alive duration for the two approaches compared to OpenWhisk.

In both experiments, accurate invocation forecasting enabled timely container reclamation, significantly improving resource efficiency.  
Under the Azure Function trace, MPC-Scheduler reduced the total number of warm containers by 34.8\% and cut keep-alive duration by 64.3\%, while IceBreaker achieved reductions of 17.4\% and 43\%, respectively -- relative to OpenWhisk’s default 10-minute keep-alive policy.
Under the synthetic bursty workload, although warm containers were more actively utilized due to high request arrival rates (resulting in less idle time), both MPC-Scheduler and IceBreaker still reduced resource usage relative to OpenWhisk. Specifically, MPC-Scheduler and IceBreaker reduced warm container usage by 19.1\% and 14.8\%, respectively, while also shortening keep-alive durations by 15.7\% and 11.3\%.

IceBreaker’s gains rely on exploiting server heterogeneity for cost-effective function placement. However, in our homogeneous testbed, its utility function loses this advantage, limiting its ability to minimize keep-alive costs and maintain warm containers under tight budgets. This limitation largely explains why IceBreaker underperforms compared to MPC-Scheduler in our experimental setting.

\textit{Overall, the observations show that the MPC-Scheduler effectively reduces resource usage and shortens function keep-alive durations, thereby lowering the overall cost of maintaining warm containers. }


\subsection{Control overhead}
\begin{figure}[H]
  \centering
  \includegraphics[width=0.5\columnwidth]{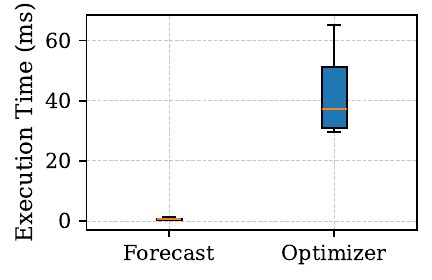}
\caption{Breakdown of execution time for each component of the MPC scheduler.}

  \label{fig:overhead}
\end{figure}

Our MPC-based scheduler adds minimal overhead, making it suitable for real-time deployment. At each control interval, it performs two tasks: forecasting future invocation rates and solving the optimization problem. As shown in Figure~\ref{fig:overhead}, the forecasting step is fast, averaging just 0.1\,ms, while the optimizer completes in 38ms on average. 

\textit{These results demonstrate that the control logic can operate at fine-grained intervals without introducing a performance bottleneck.}

\subsection{Limitation}
The effectiveness of the MPC scheduler relies on the accuracy of the invocation forecast, as these predictions guide decisions on how many requests to serve and how many function replicas to prewarm. Furthermore, as observed in our experiments, when the request arrival pattern is steady -- such as in typical Azure Functions workloads -- the benefits of MPC in reducing cold start latency become marginal.  In these scenarios, even OpenWhisk’s default behavior of keeping containers warm for up to 10 minutes is generally sufficient to handle regular traffic without triggering frequent cold starts. 


Another important consideration is MPC parameter tuning, as performance is sensitive to choices like the control horizon and objective weights. Our parameters were empirically calibrated, but this ad hoc approach is workload-specific. Future work should explore automated online tuning (e.g., Bayesian optimization, reinforcement learning, meta-learning) to adapt under dynamic workloads.

%% file: conclusion.tex
Serverless computing simplifies cloud deployment with an event-driven model that hides infrastructure complexity. Yet, cold start delays remain a major challenge for latency-sensitive workloads.

This paper introduced a novel scheduling framework based on MPC to proactively reduce cold start impact and minimize end-to-end latency. By forecasting future request arrivals, the controller jointly optimizes container prewarming and request shaping, enabling smoother load handling and improved responsiveness. We implemented our approach in Apache OpenWhisk and evaluated it on a Kubernetes-based testbed. 
Results show that MPC-Scheduler outperforms state-of-the-art approaches, achieving up to 85\% lower 90th percentile tail latency and 34\% fewer resource usages compared to the default OpenWhisk policy.

Future work includes extending this framework to heterogeneous multi-platform orchestration, enabling coordinated scheduling across hybrid or edge–cloud deployments.